\documentclass[a4paper,12pt,reqno]{amsart}

\usepackage[centertags]{amsmath}
\usepackage{amsfonts}
\usepackage{amssymb}
\usepackage{amsthm}
\usepackage{newlfont}
\usepackage{stmaryrd}
\usepackage{mathrsfs}
\usepackage{graphics}
\theoremstyle{plain}

\theoremstyle{definition}

\theoremstyle{remark}

\numberwithin{equation}{section}

\let\al=\alpha \let\be=\beta  \let\ep=\epsilon

\let\si=\sigma

\newcommand{\caD}{{\mathcal D}}

\newcommand{\caJ}{{\mathcal J}}

\newcommand{\caL}{{\mathcal L}}

\newcommand{\caP}{{\mathcal P}}

\newcommand{\opunit}{\text{1}\kern-0.22em\text{l}}

\DeclareMathAlphabet{\mathpzc}{OT1}{pzc}{m}{it}

\newcommand{\id}{\textrm{d}}

\begin{document}

\begin{center}
\noindent{\large \bf On the validity of entropy production
principles\\ for linear electrical circuits} \\

\vspace{15pt}

{\bf Stijn Bruers,\;\; Christian Maes\footnote{{\tt
http://itf.fys.kuleuven.be/\~{}christ/}}}\\
Instituut voor Theoretische Fysica, K.U.Leuven\\
\vspace{5pt} and\\ \vspace{5pt} {\bf Karel
Neto\v{c}n\'{y}\footnote{email: {\tt netocny@fzu.cz}}}\\
Institute of Physics AS CR, Prague
\end{center}

\vspace{20pt}
\footnotesize


\vspace{20pt} \footnotesize \noindent {\bf Abstract: } We discuss
the validity of close-to-equilibrium entropy production principles
in the context of linear electrical circuits. Both the minimum and
the maximum entropy production principle are understood within
dynamical fluctuation theory.  The starting point are Langevin
equations obtained by combining Kirchoff's laws with a
Johnson-Nyquist noise at each dissipative element in the circuit.
The main observation is that the fluctuation functional for time
averages, that can be read off from the path-space action, is in
first order around equilibrium given
by an entropy production rate.\\
That allows to understand beyond the schemes of irreversible
thermodynamics (1) the validity of the least dissipation, the
minimum entropy production, and the maximum entropy production
principles close to equilibrium; (2) the role of the observables'
parity under time-reversal and, in particular, the origin of
Landauer's counterexample (1975) from the fact that the
fluctuating observable there is odd under time-reversal; (3) the
critical remark of Jaynes (1980) concerning the apparent
inappropriateness of entropy production principles in
temperature-inhomogeneous circuits.

% -----------------------------------------------------------------------------------------------------------
\normalsize
\section{Introduction}

Fluctuation theory is a standard topic in equilibrium thermostatistics, and its relation to
thermodynamic variational principles is very well understood. In nonequilibrium these studies stem
from the fundamental work of Onsager and Machlup, and its various generalizations  are nowadays a
hot topic, \cite{jona,der,jor}. The existence of a link between variational  principles like that
of least dissipation and the Onsager-Machlup Langrangians has been known  for a long time.
However, recent progress in better understanding the role of time-reversal symmetry and its
breaking (cf.\ fluctuation theorems, Jarzynski identity etc., see e.g.\ Refs.~\cite{der,poincare})
allows to analyze this in a fresh and more systematic way, and to clean up some old ambiguities
and to reinterpret some of the existing formulations.
That is the motivation of the present paper.\\

A traditional way of illustrating how entropy production principles characterize the behavior of
(non)equilibrium systems goes via the study of linear electrical circuits. The reason is that they
provide physically clear and mathematically simple testing grounds for these hypotheses.  The
minimum and maximum entropy production principles have a reputation of being vague or of being at
best only sometimes valid.  In fact a well known counter example was discussed by Landauer in 1975
showing that the minimum entropy production (MinEP) principle may not be satisfied even close to
equilibrium, \cite{L1,L2}. A more general criticism was exposed by Jaynes, starting with the
remark that the MinEP does not even work when in a network the resistors are kept at different
temperatures, \cite{J}. Surprisingly, it has not stopped people from applying MinEP in a variety
of contexts and from inventing new proofs for it, see e.g.~\cite{MNS} for a critical review.

In the present paper we illustrate our new understanding of these
principles via dynamical fluctuation theory.  No longer is it
solely a matter of verifying these entropy principles but of
explaining their origin and their approximate nature.  More
mathematical details are in \cite{MN2}.  There are two particular
steps in our analysis. First we connect the discussion with more
recent work in nonequilibrium statistical mechanics, in particular
as described in \cite{MN,poincare}. We construct a Lagrangian
governing the distribution of system trajectories; the entropy
production is identified with the source term of time-reversal
breaking in that Lagrangian. Secondly, we explain how the entropy
production principles can be derived from the fluctuation theory
constructed from the Lagrangian. The basic input is the
observation that the rate function for certain stationary
dynamical fluctuations is in a direct relation with the entropy
production rate, when close to equilibrium and under specific
conditions. This yields an extension of the classical work by
Onsager and Machlup \cite{OM} which enables to systematically
generate various variational characterizations of the stationary
state. Among these we discuss the validity of both the MinEP and
of its counterpart in the maximum entropy production principle
(MaxEP). In particular we revisit Landauer's counter example and
we show why the conditions for the validity of the MinEP are not
verified.

The plan of the paper is as follows. In the next section we introduce the general framework and
formulate the main questions for which we give general answers. Afterwards we discuss some generic
linear electrical circuits to illustrate the main points of the theory.

The paper is part of a series of papers in which the entropy
production principles are revisited and extended, also in the
light of recent advances in nonequilibrium statistical mechanics,
\cite{MN1,MN2,B}.\\
The number of references on entropy production principles is enormous, mostly however within the
formalism of irreversible thermodynamics. We mention some but only a tiny fraction of them in the
course of the paper. As is well known, much of the pioneering work was of course done by Ilya
Prigogine, see Refs.~\cite{P,GM}. As a discussion of some conceptual points, we mention
Refs.~\cite{J,L3}.

\section{Set-up and general strategy}

In the present paper we explain when entropy production principles
can be expected to yield correct physical information. That will
be illustrated in the context of linear electrical circuits. Given
such a network there is often an immediate phenomenological
expression of the entropy production rate $\dot S(X,\dot X)$ as a
function of the relevant variables $X=(X_\alpha)$ (potentials,
currents), their time-derivatives $\dot X$, and in terms of the
network parameters (values of the electrical elements).  From
$\dot S(X,\dot X)$ we can still construct other entropy production
variables, e.g.\ by substituting a dynamical law for $\dot X$ one
obtains a \emph{typical} entropy production rate at state $X$. The
question is now when and why the correct physical values for the
potentials and/or the currents minimize or perhaps maximize these
entropy production rates, in whatever form. The question and part
of the answer will become more clear below. We will then give
examples of networks, write down the relevant entropy production
and show how it can serve, if at all, as a variational functional.\\

Our treatment of electrical circuits relies on the use of Kirchoff equations that themselves
assume a quasi-stationary regime of the Maxwell equations, thus forgetting about the dynamics of
the electromagnetic field.  The resulting differential equations are well known and their
stationary solutions are in the standard textbooks. Obviously our goal is not to study linear
electrical networks; in the present paper we use them merely as an example to illustrate why and
how nonequilibrium behavior can or cannot be obtained from variational principles for the entropy
production. We refer to e.g.~\cite{HR} for other and further illustrations of the use of
electrical networks in studies of nonequilibrium physics, also using stochastic methods.

For that purpose, Kirchoff's differential equations will be
embedded in stochastic equations whose averaged behavior
reproduces the deterministic equation.  Not only does that enable
the use of stochastic methods but there is also a good physical
reason to include extra noise. In accord with the
fluctuation-dissipation theorem, the thermal agitations in a
resistor are related to the distribution of the random electric
field acting upon the electrons. As a consequence, a random
voltage emerges and can be measured at the ends of the resistor
(Johnson effect).  That voltage can be described as a random
process $U^f_t$ given by the Nyquist formula:
\begin{equation}\label{nyq}
  U^f_t\,\id t = \sqrt\frac{2 R}{\be}\, \id W_t, \quad \text{or}\quad
  U^f_t = \sqrt\frac{2 R}{\be} \;\xi_t
\end{equation}
with $R$ the resistance, $W_t$ a standard Wiener process and more
formally, $\xi_t$ a standard white noise; the prefactor is of
course very small by the presence of Boltzmann's constant in
$\beta^{-1} = k_B T$, at least when compared to macroscopic
voltage values. Hence, every `real' resistor can be equivalently represented as an ideal resistor
in series with the random voltage source $U^f_t$. Using such a representation for all resistors
present, we can study fluctuations in an arbitrary electrical circuit. The resistors in the
network are the only source of fluctuations and of steady dissipation.  Apart from the transient
contributions coming from capacitances or inductances, each resistor $R$ through which a current
$I$ flows and which is kept
in thermal contact with a reservoir at temperature $\beta^{-1}$, contributes a steady term $\beta
RI^2$ to the entropy production. Having thus determined the entropy production rate $\dot S(X,\dot
X)$ for the electrical circuit, we must understand how it gives rise to a variational
principle for the variables in the network.\\
The following can be skipped at first reading and one can choose to go directly to the examples in
Section \ref{exa}.

The line of reasoning will be as follows. For a given electrical circuit we write down (first law)
the conservation of charge, that the sum of all currents equals zero at every node, and (second
law) the conservation of energy, that the sum of all potential(differences) over any loop equals
zero.  In those we take care to add with every resistor the random process $U^f_t$ for an
additional fluctuating potential difference.  The basic variables are then the potentials and the
currents satisfying linear stochastic differential equations of the generic form
\begin{equation}\label{bas} \dot X_\alpha(t) =
f_\alpha + \sum_\gamma
c_{\alpha,\gamma}\,X_\gamma(t) +
\sqrt{\frac{2}{v_\alpha}}\,\xi_\alpha(t)
\end{equation}
where the $\xi_\alpha(t)$ are mutually independent standard white
noises; the $X_\alpha$ represent the fluctuating variables
(currents and potentials that can be chosen freely); the constants
$v_\alpha, c_{\alpha,\gamma}$ are determined from Kirchoff's laws
and from the Nyquist formula \eqref{nyq} and the $f_\alpha$ is the external ``force'' (such as
from an external source or battery). We will see equation \eqref{bas} specified in
\eqref{rclang1}, \eqref{rllang}, \eqref{rrcllang2}, and \eqref{eq: RRp}.\\
 That stochastic dynamics induces a
probability distribution $\mathbf P$ on histories $\omega$, where for
each time $t$, $\omega_t = (X_\alpha(t))_\alpha$ states the values
of the potentials and of the currents. The action in $\mathbf P$ is
readily computed from It\^o-stochastic analysis:
\[
  \mathbf P(\omega) \propto \exp \Bigl[ -\int \id t \, \caL(\omega_t,\dot \omega_t) \Bigr]
\]
with Onsager-Machlup Lagrangian, formally,
\begin{eqnarray}\label{ons}
  \caL(X,\dot X) = \frac 1{4}\sum_\alpha v_\alpha \bigl( \dot X_\alpha -
f_\alpha -  \sum_\gamma c_{\alpha,\gamma} X_\gamma \bigr)^2
\end{eqnarray}
From a mathematical point of view, such expressions are justified
within the Freidlin-Wentzell theory of stochastic perturbations of
deterministic evolutions~\cite{DZ}. Observe that $v_\al = O(\be)$
in the electrical circuits and $v_\alpha\, f_\alpha^2$ is a very
high frequency for not too high temperatures; therefore the
typical trajectories are $\dot X_\alpha = f_\alpha +  \sum_\gamma
c_{\alpha,\gamma} X_\gamma$ and $\be^{-1}$ can be taken
as a perturbation parameter in the theory.\\
Each time in the examples below, we will explicitly write down that Lagrangian, see \eqref{l1},
\eqref{l2}, \eqref{l4}, and \eqref{eq: L-RRp}.\\

When applying the general model~\eqref{bas} to a particular physical problem, we always have to
satisfy a consistency condition: that the antisymmetric
 term under time-reversal in $\caL$ is the
physically correct entropy production $\dot S(X,\dot X)$, usually \emph{a priori} known from the
context. It means the entropy production must satisfy
\begin{equation}\label{star}
  \caL(\varepsilon X, -\varepsilon \dot X) - \caL(X,\dot X) =  \dot S(X,\dot X)
\end{equation}
with $(\varepsilon X)_\alpha  = \varepsilon_\alpha X_\alpha$ for
parities $\varepsilon_\alpha = \pm 1$, labelling the
(anti)symmetry under kinematical time-reversal. E.g.,
$\varepsilon_\alpha=1\,(\text{or } -1)$ if $X_\alpha$ is a voltage
(or current). In our linear electrical circuits, that is ensured
by satisfying the fluctuation-dissipation relation by
taking~\eqref{nyq} as noise terms, in combination with a suitable
choice of variables or of the level of description. The latter
point is subtle: using a too coarse grained level of description
one can easily `become blind' to some contributions to the total
entropy production. Relation~\eqref{star} will be checked in each
example below, see~\eqref{antil}, \eqref{d2}, \eqref{d4}, and \eqref{eq: L-RRp}.\\
When there is no driving (no external force nor battery nor differences in temperature,...) the
dynamics certainly reduces to that of an equilibrium system in which case the entropy production
rate \eqref{star} must be a total time derivative:
\begin{equation}\label{detbal}
\int_{t_0}^{t_1} \id t \,\dot S(X(t),\dot X(t)) = \beta \,[\,H(X(t_1)) -
H(X(t_0))\,]
\end{equation}
for some energy function $H(X)$ and some inverse temperature
$\beta$.  The equation \eqref{detbal} formulates the condition of
detailed balance.

\subsection{Transient entropy production principles}\label{sec: transient}

For the purpose of obtaining variational principles, we must look back at the Lagrangian
\eqref{ons}.  We can for example fix a history $(X_\alpha(s))$ for times $s\leq  t$ before some
fixed $t$ and ask what is the most probable immediate future. Clearly, it amounts to finding the
$(\dot X_\alpha(t))$ that minimize
$\caL(X(t),\dot X(t))$, i.e., to minimize
\begin{equation}\label{minim}
\cal D_1(X,\dot X) = \frac 1{4}\sum_\alpha v_\alpha \bigl[\dot X_\alpha^2
- 2\dot X_\al \bigl( f_\alpha + \sum_\gamma c_{\alpha,\gamma} X_\gamma \bigr)\bigr]
\end{equation}
That is traditionally called the least dissipation principle
because when all variables are even, $\varepsilon_\alpha=1$, it is
easily checked that
\begin{equation}\label{minim1}
2\cal D_1(X,\dot X) = \frac 1{2}\sum_\alpha v_\alpha \dot
X_\alpha^2 - \dot S(X,\dot X)
\end{equation}
which can be traced back to mechanical and equilibrium analogues given by Rayleigh and Onsager,
see~\cite{OM}; the expression
\begin{equation}
  \cal D(\dot X)=\sum_\alpha v_\alpha \dot X_\alpha^2
\end{equation}
is sometimes called the dissipation function. The typical behavior
$\dot X_\alpha = f_\alpha + \sum_\gamma
c_{\alpha,\gamma}\,X_\gamma$ as expected from \eqref{bas}, can
thus be characterized as the one minimizing \eqref{minim1} (still
under the condition that all $\varepsilon_\alpha = 1$).  We can
rewrite that as a (transient) maximum entropy production principle
(discussed in e.g. Ref. \cite{Ziegler}).  Indeed, minimizing
\eqref{minim1} over the $\dot X$ (for given $X$) is equivalent
with maximizing  $\dot S(X,\dot X)$ under the additional
constraint that
$\cal D(\dot X) = \dot S(X,\dot X)$.\\

Alternatively, we can fix the $\dot X_\alpha$'s in \eqref{ons} and we collect the free variable
part of \eqref{ons} in what we call $\cal D_2$; we must then find the $X_\alpha$'s minimizing
\begin{eqnarray}\label{minim2}
\cal D_2(X,\dot X) = &&\frac 1{4}\sum_\alpha v_\alpha
\bigl[ \bigl( \sum_\gamma c_{\alpha,\gamma} X_\gamma\bigr)^2\nonumber\\&&
- 2(\dot X_\alpha - f_\alpha)\sum_\gamma c_{\alpha,\gamma} X_\gamma\bigr]
\end{eqnarray}
The solution will give us the typical $X_\alpha$'s. When now all
the variables are odd, $\varepsilon_\alpha=-1$, that reduces to
minimizing
\begin{equation}\label{minim3}
2\cal D_2(X,\dot X) = \frac 1{2}\sum_\alpha v_\alpha
\bigl(\sum_\gamma c_{\alpha,\gamma} X_\gamma\bigr)^2 - \dot S(X,\dot
X)
\end{equation}
With the definition of typical (or expected) entropy production
\begin{equation}\label{expe}
\sigma(X) = \dot S(X,\dot X=f + \sum_\gamma
c_{\cdot,\gamma}\,X_\gamma) \end{equation} the expression
\eqref{minim3} can be rewritten as
\begin{equation}\label{eq: minim-odd}
  2\cal D_2(X,\dot X) = \frac {1}{2}\sigma(X) - \dot S(X,\dot X)
\end{equation}
which we have to minimize over the $X_\alpha$'s or, alternatively,
$\dot S(X,\dot X)$ has to be maximized under the constraint $\dot
S(X,\dot X) = \si(X)$.\\

Remark that if either a) all variables are even and driving forces
arbitrary, or b) all variables are odd and the forces absent,
$f_\al \equiv 0$, then both the variational principles $\caD(\dot
X)/2 - \dot S(\cdot,\dot X) = \min$ \eqref{minim1} and $\si(X)/2 -
\dot S(X,\cdot) = \min$ \eqref{eq: minim-odd} are valid and in a
sense dual. The reason is that they are both corresponding to a
relaxation to equilibrium  with Lagrangian taking the symmetric
form
\begin{equation}\label{eq: OM}
  2\caL(X,\dot X) = \frac{1}{2}\caD(\dot X) - \dot S(X,\dot X) + \frac{1}{2}\si(X)
\end{equation}
a scenario originally considered by Onsager and Machlup~\cite{OM}.
However, that structure needs a modification when a true
nonequilibrium driving is present and/or when even and odd
variables mix with each other. (There was already an example above
when  all variables were odd and a driving force was switched on.)\\

In applications, most interesting is the stationary regime for
which $\dot X = 0$. We discuss that at large in the next section.

\subsection{Stationary entropy production principles}

For the stationary regime, we obtain a variational functional by
simply putting $\dot X_\al = 0$ in the Lagrangian~\eqref{ons} to
get
\begin{equation}\label{j1}
  \caJ(X) =  \frac {1}{4}\sum_\alpha v_\alpha \bigl(f_\alpha + \sum_\gamma
  c_{\alpha,\gamma} X_\gamma\bigr)^2
\end{equation}
A more fundamental point is that $\caJ$ is the large deviation
rate function for empirical averages $\int_0^T X(t)\,\id x/T$ when
$T \uparrow\infty$, i.e.,
 \begin{equation}\label{j2}
  \mathbf P\Bigl[\frac 1{T} \int_0^T X(t)\,\id t = x \Bigr]
  \simeq \exp[ - T \caJ(x)]
\end{equation}
The mathematical theory of such large deviations was initiated by Donsker and Varadhan, see
\cite{DV,DZ}. Equation~\eqref{j2} resembles Einstein's formula for equilibrium fluctuations from
which various variational characterizations of equilibrium follow, in terms of entropy and related
potentials, see also Ref.
\cite{MN1}. Our line of thinking is similar here: a systematic way
to obtain meaningful nonequilibrium variational principles is to
consider dynamical large deviations. That point of view gets of
course more interesting when the stationary state is less
explicit, like for mesoscopic systems or for more general Markov
processes~\cite{MN2}
far from equilibrium, possibly with time-dependent driving etc.\\

\subsubsection{Even variables} When all fluctuating variables are even under
time-reversal (all $\varepsilon=+1$), then it is easy to verify
that
\begin{equation}\label{center}
  \caJ(X)=\frac 1{4}\,\sigma(X)
\end{equation}
see e.g.~\eqref{eq: OM}. Thus, in that case, we get the typical
(equilibrium) values for the $X$ by minimizing the entropy
production $\sigma(X)$, which becomes zero in equilibrium. The
most basic example is in Section \ref{rcs}.

\subsubsection{Odd variables}
Including variables that are odd under time-reversal is necessary in order to obtain a truly
nonequilibrium stationary state in the present framework. We first investigate what happens when
we have {\it only} odd variables.\\
The basic idea is always that we must minimize $\caJ(X)$ but now
it differs from the physical entropy production. That is in accord
with a general observation that for odd variables the minimum
entropy production principle does not apply. Instead, minimizing
$\caJ(X)$ can now be understood as a certain maximum entropy
production principle. The reason is that for odd variables
\begin{equation}\label{centero}
  2\caJ(X) = \frac 1{2} \sigma(X) - \dot S(X,0)
  + \frac{1}{2} \sum_\alpha v_\alpha f_\alpha^2
\end{equation}
Hence, we need to minimize
\begin{equation}\label{216}
  \frac 1{2} \sigma(X) - \dot S(X,0)
\end{equation}
In the stationary case, we have $\sigma(X) = \dot S(X,0)$.  Thence, minimizing $\caJ (X)$ amounts
to maximizing the entropy production $\sigma(X)$ under the constraint that $\sigma(X) = \dot
S(X,0)$. That repeats the discussion after~\eqref{eq: minim-odd}, but this time for the
stationary case $\dot X = 0$.\\
The required modification of the MinEP (to minimizing \eqref{216}) when dealing with odd variables
is also how Landauer's counter example~\cite{L1,L2} should be understood. The details are in
Section~\ref{rlseries}.

\subsubsection{Even and odd} The above equations (\ref{center}, \ref{centero})
have been obtained for dynamical variables that are either all
even or all odd under time-reversal. We can make that more
general. Suppose our Lagrangian includes both time-reversal even
and odd variables: $\{X_\alpha\}=\{X_i^+,X_i^-\}$, where it is
understood that the $X_i^+$ are even and that the $X_i^-$ are odd
under time-reversal. The Lagrangian \eqref{ons} now takes the form
(in obvious notation):
\begin{equation}\label{onsm}
\begin{split}
  \caL(X,\dot X)
  &= \frac{1}{4}\sum_{i+} v^+_i \bigl(\dot X^+_i-f^+_i -\sum_{j+} c_{ij}^{++}X^+_j-\sum_{j-}
  c_{ij}^{+-}X^-_j \bigr)^2
\\
  &+ \frac{1}{4}\sum_{i-} v^-_i \bigl(\dot X^-_i -f^-_i -\sum_{j+}
  c_{ij}^{-+}X^+_j-\sum_{j-} c_{ij}^{--}X^-_j \bigr)^2
\end{split}
\end{equation}
Remember that from the beginning we restrict ourselves
to external driving forces which are even
under time-reversal; for odd driving
forces already \eqref{star} needs a modification but an
analogous reasoning applies.\\

Let us first consider the entropy production rates. After some
calculation, we find for the expected entropy production
\eqref{expe}:
\begin{equation}
\begin{split}
  \sigma(X^+,X^-) &= \sum_{i+} v^+_i \bigl(f^+_i+\sum_{j+} c_{ij}^{++}X^+_j \bigr)^2
\\
  &\hspace{5mm}+ \sum_{i-} v^-_i \bigl(\sum_{j-} c_{ij}^{--}X^-_j\bigr)^2
\end{split}
\end{equation}
From this we can further construct a function of the even and a function of the odd variables. We
thus get two additional, even and odd, expected entropy production rates
$\sigma^+(X^+)$ and $\sigma^-(X^-)$ given as
\begin{equation}\label{sigmapm}
\begin{split}
  \sigma^+(X^+) &= \sigma(X^+,X^-(X^+))
\\
  \sigma^-(X^-) &= \sigma(X^+(X^-),X^-)
\end{split}
\end{equation}
where for the first (even) case we insert $X^- = X^-(X^+)$ from
solving the stationary condition
\[
 f^-_i+\sum_{j+} c_{ij}^{-+}X^+_j+\sum_{j-}
c_{ij}^{--}X^-_j =0
\] and likewise, for the second (odd) case we substitute $X^+ = X^+(X^-)$
as found from
\[
f^+_i+\sum_{j+} c_{ij}^{++}X^+_j+\sum_{j-} c_{ij}^{+-}X^-_j = 0
\]
That will be explicitly visible and done in \eqref{sigmapme}.\\

The large deviation rate function~\eqref{j1} can also be calculated:
\begin{equation*}
\begin{split}
  4\caJ(X^+,X^-) &=\sum_{i+} v^+_i \bigl[\bigl(f^+_i+\sum_{j+} c_{ij}^{++}X^+_j \bigr)^2
  + \bigl(\sum_{j-} c_{ij}^{+-}X^-_j\bigr)^2
\\
  &\hspace{5mm}+2\bigl(f^+_i+\sum_{j+} c_{ij}^{++}X^+_j\bigr)\bigl(\sum_{j-} c_{ij}^{+-}X^-_j\bigr)\bigr]
\\
  &\hspace{5mm}+ \sum_{i-} v^-_i\bigl[ \bigl(f^-_i+\sum_{j+} c_{ij}^{-+}X^+_j\bigr)^2 +\bigl(\sum_{j-}
  c_{ij}^{--}X^-_j\bigr)^2
\\
  &\hspace{5mm}+2\bigl(f^-_i+\sum_{j+} c_{ij}^{-+}X^+_j\bigr)\bigl(\sum_{j-} c_{ij}^{--}X^-_j\bigr)\bigr]
\end{split}
\end{equation*}
To simplify the structure, we make here the (nontrivial)
assumption that the even and the odd variables do not mix in
$\caJ$. Hence, we require that
\begin{equation}\label{conmep}
\caJ(X^+,X^-)=\caJ^+(X^+)+\caJ^-(X^-) \end{equation}
(i.e., the cross terms are zero), with
\begin{equation}
\begin{split}
  4\caJ^+(X^+)&=\sum_{i+} v^+_i (f^+_i+\sum_{j+} c_{ij}^{++}X^+_j)^2
\\
  &\hspace{5mm}+\sum_{i-} v^-_i (f^-_i+\sum_{j+} c_{ij}^{-+}X^+_j)^2
\end{split}
\end{equation}
and analogously for $\caJ^-(X^-)$, see below in \eqref{maxm}. That
decoupling of the even from the odd variables in the rate function
$\caJ$, implies the relation
\begin{eqnarray}\label{jplus}
4\caJ^+(X^+)=\sigma^+(X^+)
\end{eqnarray}
which is a generalization of~\eqref{center}.\\
 Remember now that we
must minimize $\caJ$ (here of the form \eqref{conmep}) to obtain
the typical stationary values. Hence, if indeed the time-symmetric
and time-antisymmetric variables in $\caJ$ decouple, then
\eqref{jplus} tells that we should minimize the expected entropy
production $\sigma^+$. We will see an
example below in Section~\ref{RRseries}.\\

There is also a MaxEP principle in the above setting. Write $\caJ^-$ as
\begin{equation}\label{maxm}
\begin{split}
  4\caJ^- &= \sum_{i+} v^+_i \bigl(\sum_{j-} c_{ij}^{+-}X^-_j\bigr)^2
  +\sum_{i-} v^-_i\bigl(\sum_{j-} c_{ij}^{--}X^-_j\bigr)^2
\\
  &\hspace{5mm}+ 2\sum_{i+} v^+_i f^+_i \sum_{j-} c_{ij}^{+-}X^-_j + 2\sum_{i-} v^-_i f^-_i
  \sum_{j-} c_{ij}^{--}X^-_j
\\
  &= \sigma^-(X^-) -2\caP(f,X^-)
\end{split}
\end{equation}
where $\caP$ can be interpreted as the power input from the
external forces $f$. Suppose we take the constraint $\sigma^- =
\caP$ meaning that the delivered work is completely dissipated and
there is no accumulation of internal energy (true indeed in the
stationary state when $\dot X =0$). Minimizing $\caJ^-(X^-)$ under
the constraint $\si^- = \caP$ is equivalent to maximizing
$\sigma^-$ under the same constraint. That MaxEP principle was
used similarly in Ref.~\cite{zupanovic}.

\section{Examples}\label{exa}

We demonstrate the above general theory on a few simple examples of linear circuits. In particular
we will see that close to equilibrium the rate function $\caJ$ can indeed be split in the even and
odd parts, and hence, depending on the choice of variables, we obtain MinEP or MaxEP principle.

\subsection{RC in series}\label{rcs}
Consider a resistance $R$ in series with a capacity $C$ and with a steady voltage source $E$.
Write $U=U_t$ for the variable potential difference over the capacitor. Kirchhoff's second law
reads
\begin{equation}\label{rclang}
  RC\dot U = E - U + U^f
\end{equation}
By inserting the white noise $\xi_t$ following~\eqref{nyq}, we are to study the Langevin equation
\begin{equation}\label{rclang1}
  \dot U_t = \frac{E - U_t}{R C} + \sqrt\frac{2}{\be R C^2}\, \xi_t
\end{equation}
There is no other free variable apart from $U$; in particular, the current
$I = C\dot U$. A standard reasoning proves the consistency of this model: with the battery removed,
$E = 0$, the dynamics is reversible with respect to the Gibbs distribution at inverse temperature
$\beta$ and with energy function $H(U) = C U^2/2$. In particular,
$\lim_{t\uparrow\infty} \langle U_t^2 \rangle = (\be C)^{-1}$, in accordance with the
equipartition theorem.\\

Heuristically, the entropy production rate $\si(U)$ as a function of the voltage $U$ on the
capacitor is simply the Joule heating in the resistor $R$:
\begin{equation}\label{veri}
  \si(U) = \be \frac{(E - U)^2}{R}
\end{equation}
Apparently, its minimizer $U^* = E$  coincides with the correct value for the stationary
voltage, verifying the MinEP principle.\\
We can understand that within our general framework. The Onsager-Machlup Lagrangian
$\caL(U,\dot U)$ of the process $U_t$ is
\begin{equation}\label{l1}
  \caL(U,\dot U) =
  \frac{\be R C^2}{4} \Bigl(\dot U -
  \frac{E - U}{RC} \Bigr)^2
\end{equation}
From $\caL$ we can derive the fluctuations of the empirical
voltage $U_T = \int_0^T U_t\,\id t\,/T$. The stationary
($T\uparrow +\infty$) fluctuations of $U_T$ are given by
\eqref{j1}-\eqref{j2}:
\begin{equation}\label{rc}
 \mathbf P[U_T \simeq u] \propto \exp [-T \caJ(u)]
 \end{equation}
with rate function
\begin{equation}\label{antil1}
  \caJ(U) = \caL(U,0) = \frac{1}{4} \si(U)
\end{equation}
The fluctuation law~\eqref{rc} thus gives a variational principle for the stationary voltage just
coinciding with the MinEP principle: the most probable value for the time-averaged voltage is
obtained by minimizing $\caJ(U) = \si(U)/4$.\\
In fact, this is just a particular example of the relation~\eqref{center}. To see that we still
have to check that our model is consistent with relation~\eqref{star}. Indeed,
\begin{equation}\label{antil}
  \dot S(U,\dot U) =
  \caL(U,-\dot U) - \caL(U,\dot U) =
  \be C\dot U (E - U)
\end{equation}
is the physical entropy production rate, and its typical value~\eqref{expe} equals
$$
 \dot S\bigl(U,\dot U = \frac{E - U}{R C}\bigr) = \si(U)
$$
verifying \eqref{veri} above. Hence, from the fluctuation point of
view, the manifest validity of the MinEP principle for this
RC-circuit is nothing but a consequence of the invariance of the
voltage (or charge) with respect to time-reversal.

\subsection{RL in series}\label{rlseries}
 If the capacity in the previous section is replaced with an
inductance $L$, the situation remarkably changes. In that case,
Kirchhoff's second law for the current $I$ becomes
$$
  R I - U^f + L\frac{\id I}{\id t} = E
$$
(where the minus sign is chosen for convenience only), or,
inserting the white noise $\xi_t$ from \eqref{nyq},
\begin{equation}\label{rllang}
  \dot I_t = \frac{E - R I_t}{L} +
   \sqrt\frac{2 R}{\be L^2}\,\xi_t
\end{equation}
That is again a linear Langevin equation, but now the fluctuations concern the current $I_t$ which
is odd under time-reversal.\\

We can try to repeat the same as in the previous section. The
Lagrangian now equals
\begin{equation}\label{l2}
  \caL(I,\dot I) = \frac{\be L^2}{4 R} \Bigl(\dot I - \frac{E - R I}{L}\Bigr)^2
\end{equation}
The construction of the fluctuation rate $\caJ(I)$ for the
empirical current $\int_0^T I_t\,\id t\,/T$ can again be done
following \eqref{j1}-\eqref{j2} with the result
\begin{equation}\label{jrl}
  \caJ(I) = \caL(I,\dot I = 0) = \frac{\be R}{4}
  \Bigl(I - \frac{E}{R}\Bigr)^2
\end{equation}
As generally true, the minimum of $\caJ$ over $I$ is given by the correct stationary value.
However, that $\caJ$ clearly differs from the physical entropy production the expected rate of
which is now
\begin{equation}
  \si(I) = \beta RI^2
\end{equation}
So while we can find the most probable current $I^* = E/R$ by minimizing $\caJ(I)$, it does not
correspond to a minimization of the entropy production.  Indeed, our RL-circuit is the classical
example first given by Landauer through which we see that the MinEP principle is not generally
valid and `not reliable' when applied to macroscopic systems, even in the linear irreversible
regime.  We can now however understand what is the real cause of that effect.\\

Similarly to \eqref{antil1}-\eqref{antil}, the variational functional $\caJ(I)$ of \eqref{jrl} for
the stationary current satisfies
\begin{equation}\label{f1}
  \caJ(I) = \frac{1}{4}\bigl[\caL(I,-\dot I) - \caL(I,\dot I)\bigr]
  \bigr|_{\dot I = \frac{E - R I}{L}} %\neq \si(I)
\end{equation}
but $\caL(I,-\dot I) - \caL(I,\dot I)$ does not longer coincide
with the variable entropy production $\dot S$. Since the current
is odd under time reversal, the latter is rather, see
\eqref{star},
\begin{equation}\label{d2}
  \dot S(I,\dot I) = \caL(-I,\dot I) - \caL(I,\dot I) =
  \be I(E - L \dot I)
\end{equation}
in accordance with the phenomenology; note that the equilibrium dynamics (i.e.\ \eqref{rllang}
with $E=0$) satisfies detailed balance with energy function $H(I)=LI^2/2$.\\
Via our fluctuation approach we thus understand the origin of the problem: the MinEP principle is
generally valid only for Markovian dynamical systems described via a collection of observables
that are symmetric under time-reversal. The Markovian property refers to the first order (in time)
of the dynamical equation and indicates that the variable in question thus satisfies an autonomous
equation.\\
Observe that we now see appear the `true' variational principle for the stationary current as was
explained under \eqref{centero}-\eqref{216}: for odd observables the above argument proposes a
different functional that replaces the entropy production $\si(I)$ and that can here be chosen as
$$
  \frac{1}{2}\si(I) - \dot S(I,0)
$$
as follows from \eqref{jrl} written in the form
\[
\caJ(I) = \frac 1{4}\sigma(I) - \frac 1{2}\dot S(I,0) +
\frac{\beta E^2}{4R}
\]

As the stationary value $I^*$ satisfies $\sigma(I^*) = S(I^*,0)$,
we can also state the above variational principle as a maximum
entropy production principle:  we must maximize  $\sigma(I)$
subject to the condition that $\dot S(I,0) =\sigma(I)$.

\subsection{RR in series}\label{RRseries}

\begin{figure}
\includegraphics{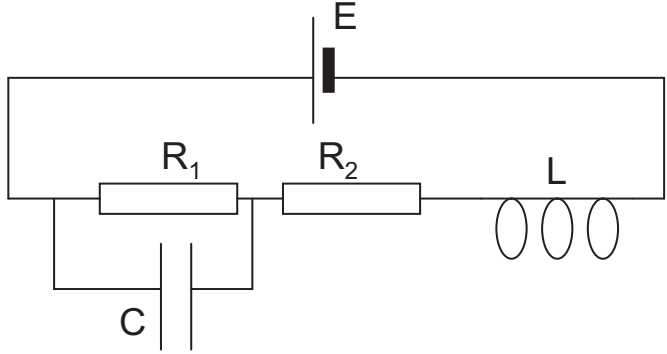}
\caption{The regularized RR-circuit.}
\label{fig}
\end{figure}

Consider an electrical circuit consisting of a battery $E$ coupled to resistors in series.
Appending an independent Nyquist random voltage source $U^f_k$ to all resistances $R_k$, we get
that the current $I$ fluctuates according to
\begin{equation}\label{sing}
  I \sum_k R_k = E + \sum_k U^f_k
\end{equation}
Here, the current follows a \emph{singular} Markov process of the form of a white noise. We can
however modify that singular dynamics so that it becomes a regular Markov process and so that the
(averaged) stationary current remains unchanged. An important addition to the discussion is to see
the effect of assigning different temperatures $\be_k$ to the individual resistors, as it was
claimed that such an extension would again yield a counter example to the MinEP, \cite{J}.\\
For regularization we choose to add an inductance in series with the resistors so that a
non-trivial transient regime can arise. However, adding only an inductance gives a too coarse
grained description, because as one can check, then \eqref{star} would not be satisfied with the
correct expression for the entropy production. That can be solved by adding a capacitance in
parallel with one of the resistors, see Fig.~\ref{fig}.  We thus have an effective RLC-circuit
with two resistors in series and one external voltage. The two independent free variables are the
potential $U$ over the first resistor and the current $I$ through the second resistor. The
variational functional \eqref{j1} or the expected entropy production
\eqref{expe} will not depend on the auxiliary inductance $L$ or on the capacitance $C$.\\

The dynamical equations are given by Kirchoff's laws:
\begin{equation}\label{rrcllang}
\begin{split}
  \dot U &= \frac{I}{C} - \frac{U}{R_1C} + \frac{U_1^f}{R_1C}
\\
  \dot I &= \frac{E - R_2I - U}{L} + \frac{U_2^f}{L}
\end{split}
\end{equation}
where always from \eqref{nyq},
$U^f_{1,2}\,\id t = \sqrt{2R_{1,2}\beta^{-1}_{1,2}}\,\id W_t$. The equilibrium dynamics
($E=0, \beta_1=\beta_2=\beta$ in \eqref{rrcllang}) can be written as
\begin{equation}\label{rrcllang2}
\begin{split}
  \dot U &=
   \frac 1{LC} \frac{\partial H}{\partial I} -
   \gamma_1 \frac{\partial H}{\partial U}
   + \sqrt{\frac{2 \gamma_1}{\beta}}\,\xi_1
\\
  \dot I &= - \frac 1{LC}\frac{\partial H}{\partial U} - \gamma_2
   \frac{\partial H}{\partial I}
  + \sqrt{\frac{2 \gamma_2}{\beta}}\,\xi_2
\end{split}
\end{equation}
for energy function $H(U,I) = (CU^2 + LI^2)/2$ and friction coefficients
$\gamma_1 = 1/(R_1C^2)$, $\gamma_2 = R_2/L^2$; the $
\xi_1$ and $\xi_2$ are independent standard white noises.  The first terms on
the right hand-side of \eqref{rrcllang2} specify a Hamiltonian
dynamics for the pair $(U,I)$, while the other terms balance the
dissipation and the random forcing.\\ %The entropy production in
The (nonequilibrium) Lagrangian is obtained from \eqref{rrcllang} as
\begin{eqnarray}\label{l4}
\caL(U,I;\dot U,\dot I) &=&\frac{\beta_1R_1C^2}{4}\Bigl(\dot U -
\frac{I}{C} + \frac{U}{R_1C}\Bigr)^2 \nonumber\\&&+
\frac{\beta_2L^2}{4R_2}\Bigl(\dot I + \frac{R_2I}{L} - \frac{E
-U}{L}\Bigr)^2 \nonumber\\&&
\end{eqnarray}
and the variable entropy production \eqref{star} is
\begin{equation}
\begin{split}\label{d4}
  \dot S(U,I;\dot U,\dot I) &= \caL(U,-I;-\dot U,\dot I) -
  \caL(U,I;\dot U,\dot I)
\\
  &= \beta_1 U (I-C\dot U) +\beta_2I(E - U - L\dot I)
\end{split}
\end{equation}
as it should. One recognizes indeed the dissipation in each resistor; $I-C\dot U$ is the current
through resistor $1$ and $E - U - L\dot I$ is the voltage at resistor $2$.  Thus, the expected
entropy production \eqref{expe} is
 \begin{equation}\label{f2}
 \si(U,I) = \beta_1\frac{U^2}{R_1} +
\beta_2 R_2 I^2 \end{equation}
On the other hand, the true variational functional \eqref{j1}-\eqref{j2} is directly obtained from
\eqref{l4}:
\begin{equation}\label{fe4}
\begin{split}
  4\caJ(U,I) &= \si(U,I)
  + \be_1 R_1 I^2 + \beta_2\frac{(E - U)^2}{R_2}
\\
  &\hspace{5mm}-2[\be_1 U I + \be_2(E - U) I]
\end{split}
\end{equation}
and differs from the entropy production because we have both an even (the potential $U$) and an
odd (the current $I$) degree of freedom; compare with \eqref{center} and \eqref{centero} valid in
the case of only even respectively only odd variables. So now we have to go to the formalism with
mixed variables as in \eqref{onsm}--\eqref{maxm}. Although $\caJ$ does not exactly split into even
and odd parts as in~\eqref{conmep}, it does approximately close to equilibrium: if
$\beta=\beta_1=\beta_2+\ep$, we have
\begin{equation}
\begin{split}
  4J(U,I) &=\beta\Bigl[ \frac{U^2}{R_1}+ \frac{(E-U)^2}{R_2} \Bigr] + \be(R_1+R_2)I^2
\\
  &\hspace{5mm}-2\be E I + O(\ep)
\end{split}
\end{equation}
which is of the form \eqref{conmep}. Since the even and the odd versions~\eqref{sigmapm} of the
entropy production rate are
\begin{align}\label{sigmapme}
  \sigma^+(U) &= \be_1\frac{U^2}{R_1} + \be_2\frac{(E - U)^2}{R_2}
  = \be \Bigl[ \frac{U^2}{R_1} + \frac{(E - U)^2}{R_2} \Bigr] + O(\ep)
\\\intertext{and}
  \sigma^-(I) &= \beta_1 R_1 I^2 + \beta_2 R_2 I^2 = \be (R_1 + R_2) I^2 + O(\ep)
\end{align}
the minimization of $\caJ$ provides us with the next two variational principles:\\
First, the stationary voltage $U^*$ is obtained from minimizing $\caJ^+(U) = \si^+(U)/4$, which is
a (generalized) MinEP principle~\eqref{jplus}.\\
Second, minimizing $\caJ^-(I) = \si^-(I) - 2\be E I$ or, equivalently, maximizing
$\si^-$ under the constraint $\sigma^-(I) = \be E I$ yields the stationary current $I^*$; this is
an example on the MaxEP principle~\eqref{maxm}.\\

An important remark is that the above derivation of the MinEP and the MaxEP principles was based
not only on the assumption that the temperature is approximately homogenous but also on the
linearity of the stochastic model(s) under consideration. They should be really taken as a
(linear) approximation around detailed balance. In particular, also the current
$I$ and the forces $U$ and $E$ should be considered of order $O(\ep)$, together with the assumed
$\be_2 = \be_1 + O(\ep)$. The above then means that the MinEP and MaxEP principles are only valid up to order
$O(\ep^2)$, i.e.\ within the linear irreversible regime; see~\cite{MN2}
for some more details.\\
In view of this remark we understand better why these principles do not carry over to
temperature-inhomogeneous circuits: in our simple circuits the temperature gradients are
\emph{redundant} thermodynamic forces in the sense that they do not generate electric currents
by themselves; they only modify the dynamics of fluctuations. Hence, these gradients yield
corrections of order $o(\ep^2)$, beyond the resolution of the MinEP/MaxEP principles. This solves
the remarks by Jaynes~\cite{J}. Apparently, the picture would get completely changed by adding
e.g.\ a thermocouple into the network.

\subsection{RR in parallel}

For the sake of completeness, we finally consider two resistors in parallel and coupled with an
external voltage source $E$. The independent variables are the currents
$I_1$ and $I_2$ through the two resistors. To have a dynamics consistent with our condition~\eqref{star}
we again need a regularization and for that we add two inductances in series with the resistances.
The resulting stochastic dynamics is
\begin{equation}\label{eq: RRp}
\begin{split}
  L_1\dot I_1 &= E-R_1I_1+U_1^f
\\
  L_2\dot I_2 &=E-R_2I_2+U_2^f
\end{split}
\end{equation}
The Lagrangian, the entropy production rate, its expectation, and the stationary variational
functional are subsequently as follows:
\begin{equation}\label{eq: L-RRp}
\begin{split}
  4\mathcal L &=\frac{\beta_1}{R_1}(L_1\dot I_1+R_1I_1-E)^2
  +\frac{\beta_2}{R_2}(L_2\dot I_2+R_2I_2-E)^2
\\
  \dot S (I_1,I_2;\dot I_1;\dot I_2)
  &= \beta_1 I_1 (E - L_1 \dot I_1) + \beta_2 I_2 (E - L_2 \dot I_2)
\\
  \sigma(I_1,I_2) &= \beta_1 R_1 I_1^2 +\beta_2 R_2 I_2^2
\\
  4\caJ(I_1,I_2) &= \sigma(I_1,I_2)-2\dot S(I_1,I_2,0,0)
  + \be_1 \frac{E^2}{R_1}+\be_2\frac{E^2}{R_2}
\end{split}
\end{equation}
If $\beta_1=\beta_2+\ep$, minimizing the rate function $\caJ$ results in minimizing
$\si(I_1,I_2) - 2\be E(I_1 + I_2)$ up to order $\ep$. Note that this splits into two independent
variational problems for $I_1$ respectively $I_2$ which comes as no surprise: the two currents are
entirely dynamically decoupled. Yet, one can again formulate a single MaxEP principle in this
case: the stationary currents are obtained by maximizing the expected entropy production rate
$\sigma$ under the constraint $\si(I_1,I_2) = \be E(I_1 + I_2)$,
the right hand side being just the total power input. This formulation is equivalent with the
MaxEP principle in~\cite{zupanovic}. Analogous remarks as in Section~\ref{rlseries} apply here.

\section{Conclusions}

Fluctuation theory naturally identifies the specific structure of dynamical fluctuations as the
underlying reason for the (variational) entropy production principles as well as their validity
conditions and limitations; just as the equilibrium fluctuation theory explains the maximum
entropy principle and its derivations. It also provides a natural and systematic way how to search
for new variational  principles beyond trial-and-error methods, not available within
pure  thermodynamics.\\

We summarize our findings within that more general perspective:\\

\begin{itemize}
\item Both MinEP and MaxEP principles have various forms (for
hydrodynamic models in local equilibrium, for discrete networks,
for various types of Markov processes,...), but there is no
essential difference between them up to the crucial restriction to
the close-to-equilibrium regime. All attempts for their exact
justification beyond the linear regime are highly doubtful. In
particular, thermodynamics itself has little to say here about
possible generalizations except via some trial-and-error methods.
We expect that dynamical fluctuation theory will present a more
systematic avenue for evaluating nonequilibrium behavior via
variational methods.

\item
There is no fundamental difference between the nature and validity of the MinEP and the MaxEP
principles. Which one is to be used depends on the  choice of thermodynamic variables, whether
they are symmetric or antisymmetric with respect to time-reversal. That clear observation is a
first concrete result that the fluctuation/statistical approach provides.

\item  The set-up in the pioneering works of Prigogine~\cite{P}
concerning MinEP typically refers  to some canonical thermodynamic structure. Yet, it is useful to
broaden the view on these variational principles; the MinEP principle is valid for mesoscopic
systems (Markovian, both with discrete and with continuous state space) where one does not {\it a
priori} recognize some canonical structure as is usually written in terms of forces and currents.
\end{itemize}

\vspace{5mm}
\noindent{\bf Acknowledgment}\\
K.~N.\ is grateful to the Instituut voor Theoretische Fysica, K.~U.~Leuven for kind hospitality.
He also acknowledges the support from the Grant Agency of the Czech Republic (Grant no.
202/07/0404). C.~M.\ benefits from the Belgian Interuniversity Attraction Poles Programme P6/02.

% --- References ---
\bibliographystyle{plain}

\end{document}